\documentclass[onecolumn, 11pt]{IEEEtran}
\usepackage{amsmath}
\usepackage{amsfonts}
\usepackage{cases}

\usepackage{amssymb}
\usepackage{graphicx}
\usepackage{color}

\newtheorem{Theorem}{Theorem}
\newtheorem{Corollary}{Corollary}
\newtheorem{Definition}{Definition}

\newtheorem{Remark}{Remark}
\newtheorem{Lemma}{Lemma}

\newtheorem{Conjecture}{Conjecture}

\newcommand{\F}{\mathbb{F}}
\newcommand{\Z}{\mathbb{Z}}

\begin{document}

\title{A Note on a Conjecture for Balanced Elementary Symmetric Boolean Functions}


\author{Wei~Su,~
        Xiaohu~Tang, 
        and~Alexander~Pott
\thanks{W. Su is with the Institute of Mobile Communications, Southwest
Jiaotong University,
Chengdu, 610031, China, and also with the Institute for Algebra and Geometry (IAG),
 Otto-von-Guericke University Magdeburg, D-39106 Magdeburg, Germany (e-mail: weisu0109@
 googlemail.com).}
\thanks{X. Tang is with the Institute of Mobile Communications, Southwest
Jiaotong University, Chengdu, 610031, China (e-mail: xhutang@
ieee.org).}
\thanks{A. Pott is with the Institute for Algebra and Geometry (IAG), Otto-von-Guericke University Magdeburg,
D-39106 Magdeburg, Germany (e-mail:
alexander.pott@ovgu.de).}

}

\maketitle

\begin{abstract}
In 2008, Cusick {\it et al.} conjectured that certain elementary symmetric
Boolean functions of the form $\sigma_{2^{t+1}l-1, 2^t}$ are the
only nonlinear balanced ones, where $t$, $l$ are any positive integers, and $\sigma_{n,d}=\bigoplus_{1\le i_1<\cdots <i_d\le n}x_{i_1}x_{i_2}\cdots x_{i_d}$ for positive integers $n$, $1\le d\le n$. In this note, by analyzing the weight
of $\sigma_{n, 2^t}$ and $\sigma_{n, d}$, we prove that ${\rm
wt}(\sigma_{n, d})<2^{n-1}$ holds in most cases, and so does the conjecture.
According to the remainder of modulo 4, we also consider the weight of
$\sigma_{n, d}$ from two aspects: $n\equiv 3({\rm mod\ }4)$ and
$n\not\equiv 3({\rm mod\ }4)$. Thus, we can simplify the conjecture.
In particular, our results cover the most known results. 
In order
to fully solve the conjecture, we also consider the weight of
$\sigma_{n, 2^t+2^s}$ and give some experiment results on it.
\end{abstract}



\section{Introduction}

Boolean functions are frequently used in the design of stream
ciphers, block ciphers and hash functions. One of the most vital
roles in cryptography of Boolean functions is to be used as filter
and combination generators of stream ciphers based on linear
feedback shift registers (LFSRs). Among all the Boolean functions,
symmetric Boolean functions are an interesting subclass for their
advantage in both implementation complexity and storage space.

Symmetric Boolean functions are characterized by the fact that their
outputs only depend on the Hamming weights of their inputs. These
functions can be represented in a very compact way both for their
algebraic normal forms and for their value vectors,
which considerably reduces the amount of memory
required for storing the function and is of great interest in
software applications. Elementary symmetric
Boolean function is the basic unit composing of symmetric Boolean functions.
Some cryptographically significant properties
of (elmentary) symmetric Boolean functions have been studied in
\cite{Savicky94}-\cite{Ou12_eprint}.

Balancedness is the compulsory  property for a Boolean function,
since our cryptographic primitives is necessary to be be unbiased in
output. Recently, there are some results about the balancedness of
elementary symmetric Boolean function $\sigma_{n, d}$ :
\begin{eqnarray*}
\sigma_{n,d}=\bigoplus_{1\le i_1<i_2<\cdots <i_d\le n}x_{i_1}x_{i_2}\cdots x_{i_d},
\end{eqnarray*}
for $2\le d\le n$.

In \cite{Cusick08_IT}, Cusick {\it et al.} proved that
$\sigma_{2^{t+1}l-1, 2^t}$ is balanced if $t$ and $l$ are positive integers (Theorem 3).
Further, they presented the following conjecture.
\begin{Conjecture}\label{Conjecture_08} There are no
nonlinear balanced elementary symmetric Boolean functions except for
$\sigma_{2^{t+1}l-1, 2^t}$,
where
$t$ and $l$ are any positive integers.
\end{Conjecture}

Towards this conjecture, some results have been  obtained in
\cite{Cusick09_JMC}-\cite{Ou12_eprint}.
\begin{enumerate}

\item If $d>1$ is odd, then $\sigma_{n, d}$ is not balanced (Lemma 3.11, \cite{Cusick09_JMC});

\item If $d=2^t$, then $\sigma_{n,
d}$ is balanced if and only if $n$ has the form of $n=2^{t+1}l-1$,
where $t$ and $l$ are any positive integers 
(Corollary 3.10 and Lemmas 3.1, 3.17, \cite{Cusick09_JMC});

\item Let $n=2^{t+1}l-1$ for some positive integers $t$, $l$. If $d$ is even and
$2^t<d<2^{t+1}$, then $\sigma_{n, d}$ is not balanced 
(Corollary 3.10 and Lemmas 3.1, 3.13, \cite{Cusick09_JMC});

\item
Let $n=2^{t+2}l+r-1$, where $t$, $l>0$ and $0\le r\le 2^{t+1}$. If
$d$ is even and $2^t<d\le 2^{t+1}-2$, then $\sigma_{n, d}$ is not
balanced (Corollary 3.10 and Lemmas 3.1, 3.18, \cite{Cusick09_JMC}).
Error correctly, the authors claimed in \cite{Cusick09_JMC} that this result
holds for $0\le r<2^{t+1}+2^t$, but the proof only work for $0\le r\le 2^{t+1}$. It
 will be explained in details in Remark
\ref{Remark Cusick09 3.18};


\item
If $n=2^{t+1}l-1$, $l$ odd and $2^{t+1}\not |d$,
$\sigma_{n,d}$ is balanced if and only if $d=2^k$, $1\le k\le t$
(Theorems 1, 2, 3 \cite{Gao11_IT});

\item Conjecture \ref{Conjecture_08} holds for sufficiently large $n$. In particular, if $d$
is not a power of two, then $\sigma_{n, d}$ is not balanced for sufficiently large $n$
(Remark 3 \cite{Castro11});

\item Let $r=\lfloor {\rm log}_2 d\rfloor+1$. For any $n$,
$n>-2({\rm log}_2\cos(\frac{\pi}{2^r}))^{-1}$, all these nonlinear
balanced elementary symmetric Boolean functions are of the form
$\sigma_{2^{t+1}l-1, 2^t}$, where $t$ and $l$ are any positive
integers (Theorem 3 \cite{Guo12_eprint}). This implies that
Conjecture \ref{Conjecture_08} is true for large enough
$n$;

\item Let $d=2^{t+w}(1+2^1+\cdots+2^s)$ and $n=2^{t+w+1}(1+2^1+\cdots+2^s)+2^tq+m$,
$m\in\{-1, 0\}$. If the nonnegative integers $t$, $w$, $s$, $q$ satisfy
certain conditions, then $\sigma_{n, d}$ is not balanced (see Theorems
1-4 in \cite{Ou12_eprint} for more details).
\end{enumerate}

%
%
%
%
%
%

In this note, we first consider the weight of $\sigma_{n,2^t}$. By
applying the relationship between $\sigma_{n, d}$ and $\sigma_{n,
2^t}$, we prove that ${\rm wt}(\sigma_{n, d})<2^{n-1}$ holds in most
cases. 
Especially, these results cover
the results given in \cite{Cusick09_JMC}.

Next according to the remainder of modulo 4, we consider the weight
of $\sigma_{n, d}$ from two aspects: $n\equiv 3({\rm mod\ }4)$ and
$n\not\equiv 3({\rm mod\ }4)$. Most notably, our results cover the
results in \cite{Gao11_IT}. 
Further,
we prove that if $n=2^{t+1}l-1$, $l\ge 3$ odd and $2^{t+1}|d$,
$\sigma_{n,d}$ is not balanced for ${\rm wt}(d)=1$ or $2d\not\preceq
n$, which is not available in \cite{Gao11_IT}. For $n\not\equiv
3({\rm mod\ }4)$, we get some similar results as that of $n\equiv
3({\rm mod\ }4)$:
\begin{enumerate}

\item If $n\not\equiv 3({\rm mod\ }4)$, then
$\sigma_{n, 2^s}$ is not balanced, for any $1\le s\le \lfloor{\rm
log}_2n\rfloor$;

\item If $n\not\equiv 3({\rm mod\ }4)$, then $n$ can be written as
 $n=2^{t+1}l+r$, where $l\ge 1$ is odd, $t\ge 1$, and $r\in\{0, 1,
 2\}$.
Let $2\le d=2^{t+1}d'+d''\le n$ with ${\rm wt}(d)\ge 2$, $d'\ge 0$
and $0\le d''<2^{t+1}$. Then, ${\rm wt}(\sigma_{n,d})<2^{n-1}$ if
one of the following conditions holds: a) $l=1$; b) $l\ge 3$,
$d''=0$, and $d'\not\preceq \frac{l-1}{2}$; c) $l\ge 3$, $d''>0$,
and ($d'\not\preceq \frac{l-1}{2}$ or $d''\ne 2^t$).
\end{enumerate}
Thus, Conjecture \ref{Conjecture_08} can be simplified as follows.

\begin{Conjecture}\label{Our Conjecture mod 4}
Let $l\ge 3$ be odd, $t\ge 1$, $n=2^{t+1}l+r$, $r=-1$, $0$, $1$,
$2$. The elementary symmetric Boolean function $\sigma_{n, d}$  is
not balanced in the following cases:
\begin{enumerate}
\item $d=2^{t+1}d'$,
$\textrm{wt}(d')\ge 2$ and $2\le d'\preceq \frac{l-1}{2}$ for
$r=-1$, $0$, $1$, $2$;
\item $d=2^{t+1}d'+2^t$,
$1\le d'\preceq \frac{l-1}{2}$ for $r=0$, $1$, $2$.
\end{enumerate}
\end{Conjecture}
Therefore, to show that Conjecture \ref{Conjecture_08} is true, it
suffices to prove Conjecture \ref{Our Conjecture mod 4}.


In \cite{Castro11} and \cite{Guo12_eprint},
the results for Conjecture \ref{Conjecture_08} hold
when $n$ is large enough. And the
conclusions in \cite{Ou12_eprint} are only for very special $n$ and
$d$. Compared with those results, our results are different.


This note is organized as follows. Section II introduces the
notation and the related results about Boolean functions and
symmetric Boolean functions. In Section III, we give our main
results about the weight of $\sigma_{n, 2^t}$ and $\sigma_{n, d}$.
We prove that ${\rm wt}(\sigma_{n, d})<2^{n-1}$ holds in most cases.
In Section IV, we discuss the weight of $\sigma_{n, d}$ depending whether
$n\equiv 3({\rm mod\ }4)$ or $n\not\equiv 3({\rm mod\
}4)$.
And then Conjecture \ref{Conjecture_08} can be simplified as
Conjecture \ref{Our Conjecture mod 4}. In order to fully solve the
conjecture, we also consider the weight of $\sigma_{n, 2^t+2^s}$ and
give some experiment results on ${\rm wt}(\sigma_{n, 2^t+2^s})$ in
Section V.

\section{Preliminaries}

Throughout this note, let $\F_2$ be the finite field with two
elements, $n>0$ be a positive integer, and $\F_2^n$ be the
$n$-dimensional vector space over $\F_2$. To avoid confusion, we
denote the sum over $\Z$ by $+$, and the sum over $\F_2$ by
$\oplus$.

We first recall some necessary definitions and results about Boolean
functions and symmetric Boolean functions.

\subsection{Boolean Functions}

Let $\mathcal{B}_n$ be the set of all maps from $\F_2^n$ to $\F_2$.
Such a map is called an $n$-variable Boolean function. The {\it
support} of a Boolean function $f\in \mathcal{B}_n$ is defined as
$supp(f)=\{x \in \F_2^n\,|\, f(x) = 1\}$. The {\it Hamming weight}
$\textrm{wt}(f)$ of $f$ is the cardinality of $supp(f)$, i.e.,
$\textrm{wt}(f)=|supp(f)|$. The {\it Hamming weight} of a binary vector
$u=(u_1, u_2, \cdots ,u_n)\in \F_2^n$, is defined by $\textrm{wt}(u)=\sum_{i=1}^nu_i$.
We say that an $n$-variable Boolean function $f$ is {\it balanced}
if $\textrm{wt}(f)=2^{n-1}$.

Each Boolean function $f(x_1,\cdots,x_n)$ has a unique
representation by a multivariate polynomial over $\F_2$, called the
{\it algebraic normal form (ANF)}:
$$f(x_1,\cdots,x_n)=
\bigoplus\limits_{u=(u_1,u_2,\cdots,u_n)\in
\F_2^n}f_u\prod\limits_{i=1}^nx_i^{u_i},\ \ \  f_u\in \F_2.$$ 
The {\it algebraic degree} of $f$, denoted by $deg(f)$, is the
maximal value of ${\rm wt}(u)$ such that $f_u\ne 0$. A Boolean
function is called {\it affine} if it has degree at most 1. Note
that any nonconstant affine function is balanced.

\subsection{Symmetric Boolean Functions}

\begin{Definition}
A Boolean function $f$ is said to be \textit{symmetric} if
$$f(x_1, \cdots, x_n)=f(x_{\tau(1)}, \cdots, x_{\tau(n)}),$$
for any permutation $\tau$ of $\{1, 2, \cdots, n\}$.
\end{Definition}

\vspace{1mm}

Denote by $\mathcal{SB}_n$ the set of all $n$-variable symmetric
Boolean functions. The definition implies that a symmetric Boolean
function $f$ takes the same value for all the vectors with the same weight.
Therefore every $f\in \mathcal{SB}_n$ can be simply represented by a vector
$$v_f=(v_f(0), v_f(1), \cdots, v_f(n))\in\F_2^{n+1},$$ where
the component $v_f(i)=f(x)$ with $\textrm{wt}(x)=i$. The vector $v_f$ is
called the \textit{simplified value vector of $f$}.

\begin{Definition}\label{Def_Element}
For positive integers $n$ and $d$, $1\le d\le n$, the {\it
elementary symmetric Boolean function $\sigma_{n,d}$} is defined as
\begin{eqnarray*}
\sigma_{n,d}=\bigoplus_{1\le i_1<i_2<\cdots i_d\le n}x_{i_1}x_{i_2}\cdots x_{i_d}.
\end{eqnarray*}
\end{Definition}

\vspace{1mm}

Based on the elementary symmetric Boolean functions,
the algebraic normal form of $f\in \mathcal{SB}_n$ can be simplified
as follows:
$$f(x_1, \cdots, x_n)
=\bigoplus\limits_{i=0}^n\lambda_f(i)\sigma_{n,i},\ \ \
\lambda_f(i)\in\F_2.$$ The coefficients vector
$\lambda_f=(\lambda_f(0), \lambda_f(1), \cdots, \lambda_f(n))$ is
called the \textit{simplified ANF vector of $f$}.

\vspace{1mm}

Let $n$ and $m$ be two positive integers with their 2-adic expansions
$n=n_{k-1}2^{k-1}+\cdots n_12+n_0$
and
$m=m_{k-1}2^{k-1}+\cdots m_12+m_0$ respectively.
We say that $m\preceq n$ if $m_i\preceq n_i$ for all $0\le i< k$, and otherwise $m\not\preceq n$.

\begin{Lemma}(Lucas' formula)\label{Lemma Lucas' theorem}
For non-negative integers $n$ and $m$, the following congruence relation holds
\begin{eqnarray*}
{n\choose m}\equiv \prod_{i=0}^{k-1}{n_i\choose m_i} (\bmod\ 2).
\end{eqnarray*}
Then, ${n\choose m}\equiv 1 (\bmod\ 2)$ if and only if $m\preceq n$.
\end{Lemma}

\begin{Lemma}(\cite{Canteaut05_IT})\label{Lemma v_f and lambda_f}
Let $f\in
\mathcal{SB}_n$. Its simplified value vector $v_f$ and
simplified ANF vector $\lambda_f$ are related by
$$v_f(i)=\bigoplus\limits_{k\preceq i}\lambda_f(k)\ {\rm and\ }
\lambda_f(i)=\bigoplus\limits_{k\preceq i}v_f(k),\ \ \ \ \forall\
i\in\{0,1, \cdots, n\}.$$
\end{Lemma}

By Lemma  \ref{Lemma v_f and lambda_f}, we have
\begin{eqnarray}\label{eqn. balanced 0}
v_{\sigma_{n,d}}(i)=1~\textrm{iff} ~d\preceq i.
\end{eqnarray}
Thus, the weight of elementary symmetric Boolean function
$\sigma_{n, d}$ is \begin{eqnarray}\label{eqn. weight ESB_n} {\rm
wt}(\sigma_{n, d})=\sum_{i=0}^n{n\choose i}v_{\sigma_{n,d}}(i)=\sum_{d\preceq i}{n\choose i}.
\end{eqnarray}

\section{Our main Results}

In this section, we obtain our main results about the weight of
$\sigma_{n,2^t}$ and $\sigma_{n, d}$. We first consider the weight
of $\sigma_{n,2^t}$. Next analyzing the weight of $\sigma_{n, d}$,
we prove that ${\rm
wt}(\sigma_{n, d})<2^{n-1}$ holds in most cases. 
Most notably, we can easily
interpret the results in \cite{Cusick08_IT, Cusick09_JMC} by using
these results.

Let $n$ and $L$ be two positive integers with $1\le L\le n$. For $0\le i\le L-1$, denote
\begin{eqnarray*}A_n^{L}(i)=
\sum\limits_{0\le j\le n, j\equiv i({\rm mod\ }L)}{n\choose
j}.\end{eqnarray*} Since $2^t\preceq i$ iff $i=2^{t+1}i'+2^t+q$ for
some integers $i'\ge 0$ and $q$ with $0\le q\le 2^t-1$, we have
$i\equiv 2^t+q({\rm mod\ }2^{t+1})$. It follows from \eqref{eqn.
balanced 0} and \eqref{eqn. weight ESB_n} that
\begin{eqnarray}\label{eqn. wt 2^t}{\rm wt}(\sigma_{n,2^t})=
A_n^{2^{t+1}}(2^t)+A_n^{2^{t+1}}(2^t+1)+\cdots+A_n^{2^{t+1}}(2^{t+1}-1).
\end{eqnarray}
There is an equation about $A_n^{2^p}(i)$ given by Canteaut and Videau in
\cite{Canteaut05_IT}.

\begin{Lemma}(\cite{Canteaut05_IT})\label{Lemma sum_i mod}
 For positive integers $n$, $p$, $i$, we have
$$A_n^{2^p}(i)
=2^{n-p}+2^{1-p}\sum\limits_{j=1}^{2^{p-1}-1}
(2\cos\frac{j\pi}{2^p})^n\cos\frac{j(n-2i)\pi}{2^p}.$$
\end{Lemma}

\vspace{1mm}

The following lemma will be very useful for our discussion on that the weight of
$\sigma_{n, 2^t}$ is greater than, less than or equal to $2^{n-1}$.

\begin{Lemma}(\cite{Cusick09_JMC})\label{Lemma sum_T}
Let $t$ and $r$ be two positive integers. Suppose that $a_1\gneq a_3\ge a_5\ge
\cdots \ge a_J$, with $J =2K+1$, are nonnegative integers. Define
the sum
\begin{eqnarray*}
T=\sum\limits_{1\le j\le J, j\
\mathrm{odd}}a_j\sin\frac{jr\pi}{2^{t+1}}.
\end{eqnarray*} Then $T$ has the
same sign as $\sin\frac{r\pi}{2^{t+1}}$.
\end{Lemma}

With all the above preparation, we can consider the weight of $\sigma_{n, 2^t}$.
Consequently, we obtain the following results.

\begin{Theorem}\label{Thm. 2^t weight} Let $t$ be a positive integer and $d=2^t$.
For any positive integer $n\ge d$, $n$ can be written as
$n=2^{t+2}l+r$ for some integers $l\ge 0$ and $0\le r\le 2^{t+2}-1$.
Then we have
$$\begin{array}{c}{\rm wt}(\sigma_{n,d})\left\{\begin{array}{lll}
<2^{n-1},\ \ &{\rm if\ } 0\le r\le 2^{t+1}-2,\\
=2^{n-1},\ \ &{\rm if\ } r=2^{t+1}-1\ {\rm or\ } 2^{t+2}-1,\\
>2^{n-1},\ \ &{\rm if\ } 2^{t+1}\le r\le 2^{t+2}-2.
\end{array}\right.\end{array}$$
\end{Theorem}

\vspace{1mm}

{\bf Proof}: Applying Lemma \ref{Lemma sum_i mod} to \eqref{eqn. wt 2^t} in place of $p=t+1$, one has
\begin{eqnarray}{\rm wt}(\sigma_{n,2^t})&=&
A_n^{2^{t+1}}(2^t)+A_n^{2^{t+1}}(2^t+1)+\cdots+A_n^{2^{t+1}}(2^{t+1}-1)\nonumber\\
&=&\sum\limits_{i=2^t}^{2^{t+1}-1}[2^{n-(t+1)}+2^{-t}\sum\limits_{j=1}^{2^t-1}
(2\cos\frac{j\pi}{2^{t+1}})^n\cos\frac{j(n-2i)\pi}{2^{t+1}}]\nonumber\\
&=&2^{n-1}+2^{n-t}\sum\limits_{j=1}^{2^t-1}
(\cos\frac{j\pi}{2^{t+1}})^n
\sum\limits_{i=2^t}^{2^{t+1}-1}\cos\frac{j(n-2i)\pi}{2^{t+1}}.\nonumber
\end{eqnarray}
From the formula in \cite{Hansen75}:
\begin{eqnarray}\label{eqn. cos(sx+y)}
\sum\limits_{s=0}^N\cos(sx+y)=\csc\frac{x}{2}\cos(\frac{Nx}{2}+y)\sin\frac{(N+1)x}{2},
\end{eqnarray}
one gets
\begin{eqnarray*}\sum\limits_{i=2^t}^{2^{t+1}-1}\cos\frac{j(n-2i)\pi}{2^{t+1}}
&=&\sum\limits_{i=0}^{2^t-1}\cos\frac{j(n-2i-2^{t+1})\pi}{2^{t+1}}\\
&=&(-1)^j\sum\limits_{i=0}^{2^t-1}\cos\frac{j(n-2i)\pi}{2^{t+1}}\\
&=&(-1)^j\sum\limits_{i=0}^{2^t-1}\cos\frac{j(2^{t+2}l+r-2i)\pi}{2^{t+1}}\\
&=&(-1)^j\sum\limits_{i=0}^{2^t-1}\cos\frac{j(r-2i)\pi}{2^{t+1}}\\
&=&(-1)^j\csc(-\frac{j\pi}{2^{t+1}})\cos(-\frac{(2^t-1)j\pi}{2^{t+1}}+\frac{jr\pi}{2^{t+1}})
\sin(-\frac{2^tj\pi}{2^{t+1}})\\
&=&(-1)^j\csc\frac{j\pi}{2^{t+1}}\cos(-\frac{j\pi}{2}+\frac{j(r+1)\pi}{2^{t+1}})
\sin\frac{j\pi}{2}\\
&=&\left\{\begin{array}{lll}
0,\ \ &{\rm if\ } j\ {\rm is\ even},\\
-\csc\frac{j\pi}{2^{t+1}}\sin\frac{j(r+1)\pi}{2^{t+1}}, \ \ &{\rm
if\ } j\ {\rm is\ odd}.
\end{array}\right.
\end{eqnarray*}
Thus,
\begin{eqnarray*}
{\rm wt}(\sigma_{n,2^t})
=2^{n-1}-2^{n-t}\sum\limits_{1\le j\le 2^t-1, j\ \mathrm{odd}}
(\cos\frac{j\pi}{2^{t+1}})^n
\csc\frac{j\pi}{2^{t+1}}\sin\frac{j(r+1)\pi}{2^{t+1}}.\nonumber
\end{eqnarray*}
Let $a_j=(\cos\frac{j\pi}{2^{t+1}})^n \csc\frac{j\pi}{2^{t+1}}$,
$1\le j\le 2^t-1$. Then $a_1>a_2>\cdots>a_{2^t-1}>0$. Denote
$$T=\sum\limits_{1\le j\le 2^t-1, j\
\mathrm{odd}} (\cos\frac{j\pi}{2^{t+1}})^n
\csc\frac{j\pi}{2^{t+1}}\sin\frac{j(r+1)\pi}{2^{t+1}}=\sum\limits_{1\le
j\le 2^t-1, j\ \mathrm{odd}} a_j\sin\frac{j(r+1)\pi}{2^{t+1}}.$$
 By Lemma
\ref{Lemma sum_T}, 
$T$
has
the same sign as $\sin\frac{(r+1)\pi}{2^{t+1}}$. Since
$$\begin{array}{c}\sin\frac{(r+1)\pi}{2^{t+1}}\left\{\begin{array}{lll}
>0,\ \ &{\rm if\ } 0\le r\le 2^{t+1}-2,\\
=0,\ \ &{\rm if\ } r=2^{t+1}-1\ {\rm or\ } 2^{t+2}-1,\\
<0,\ \ &{\rm if\ } 2^{t+1}\le r\le 2^{t+2}-2,
\end{array}\right.\end{array}$$
and ${\rm wt}(\sigma_{n,2^t})
=2^{n-1}-2^{n-t}T$.
The result holds. \hfill$\Box$

\vspace{2mm}

By Theorem \ref{Thm. 2^t weight}, the  following corollary in
\cite{Cusick09_JMC} is obvious.
\begin{Corollary}\label{Cor. d=2^t}(\cite{Cusick09_JMC})
Let $n$ and $t$ be two positive integers with $2^t\le n$. Then
$\sigma_{n,2^t}$ is balanced if and only if $n$ can be written as
$n=2^{t+1}l-1$ for some positive integer $l$.
\end{Corollary}

Let $i=\sum_{k=0}^{m}i_k2^k$ and $j=\sum_{k=0}^{m}j_k2^k$ with $i_k,
j_k\in\F_2$. Define operation $\vee$:
$$i\vee
j=\sum_{k=0}^{m}\max\{i_k, j_k\}2^k.$$ Denote $\lfloor x\rfloor$ as
the largest integer less than or equal to $x$. By equation (\ref{eqn. balanced 0}), we have
the following Lemma.

\begin{Lemma}\cite{Braeken05}\label{Lemma ESB Decomposition} Let $m=\lfloor{\rm
log}_2n\rfloor$, $i=\sum_{k=0}^{m}i_k2^k$ and
$j=\sum_{k=0}^{m}j_k2^k$ with $i_k, j_k\in\F_2$. Then we have
\begin{enumerate}
\item
$\sigma_{n, j}=\sigma_{n, 1}^{j_0}\sigma_{n, 2}^{j_1}\sigma_{n,
2^2}^{j_2} \cdots \sigma_{n, 2^m}^{j_m}$;

\item $\sigma_{n, i}\sigma_{n, j}=\sigma_{n, i\vee j}$.
\end{enumerate}
\end{Lemma}

\vspace{1mm}

Based on  Theorem \ref{Thm. 2^t weight} and  Lemma \ref{Lemma ESB
Decomposition}, we can get the following three corollaries about
${\rm wt}(\sigma_{n, d})$ with ${\rm wt}(d)\ge 2$.

\begin{Corollary}\label{Cor. Weight} Let $n$, $d$ be two positive integers with
$d=2^{d_1}+2^{d_2}+\cdots +2^{d_s}$, $s\ge 2$ and $0\le
d_1<d_2<\cdots<d_s\le \lfloor{\rm log}_2n\rfloor$. If there exists $1\le i\le s$, such that
${\rm wt}(\sigma_{n, 2^{d_i}})\le 2^{n-1}$. Then ${\rm
wt}(\sigma_{n, d})<2^{n-1}$.
\end{Corollary}

{\bf Proof}: Using Lemma \ref{Lemma ESB Decomposition}, we have
$$\sigma_{n, d}=\sigma_{n, 2^{d_1}}\sigma_{n, 2^{d_2}}
\cdots \sigma_{n, 2^{d_s}}.$$ For any $0\le k\le n$, one has that
$$v_{\sigma_{n, d}}(k)=1\Longleftrightarrow v_{\sigma_{n,
2^{d_j}}}(k)=1,\ \ \forall 1\le j\le s,$$ which implies
$supp(\sigma_{n, d})\subseteq supp(\sigma_{n,2^{d_j}})$ for any $1\le j\le s$. But,
$v_{\sigma_{n, 2^{d_i}}}(2^{d_i})=1$, $v_{\sigma_{n,
2^{d_j}}}(2^{d_i})=0$ for $j\ne i$, and then  $v_{\sigma_{n,
d}}(2^{d_i})=0$.  Thus, $supp(\sigma_{n, d})\subset
supp(\sigma_{n,2^{d_i}})$ and
$${\rm wt}(\sigma_{n,d})<{\rm
wt}(\sigma_{n,2^{d_i}})\le 2^{n-1}.$$
 \hfill$\Box$

\begin{Corollary}\label{Cor. odd degree}
Let $n$, $d$ be two positive integers with $d>1$ being odd. Then ${\rm
wt}(\sigma_{n, d})<2^{n-1}$.
\end{Corollary}

{\bf Proof}: If $d>1$ is odd, then $d=1+d'$, where $d'\ge 2$ is
even. By applying Lemma \ref{Lemma ESB Decomposition}-2), one has
$\sigma_{n, d}=\sigma_{n,1}\sigma_{n, d'}$. Clearly,  ${\rm
wt}(\sigma_{n,1})=2^{n-1}$ since $\sigma_{n,1}$ is a linear function
and then is balanced. If follows from Corollary \ref{Cor. Weight}
that ${\rm wt}(\sigma_{n, d})<2^{n-1}$.\hfill$\Box$

\vspace{1mm}

\begin{Corollary}\label{Cor. even degree} 
Let $n$, $d$ be two positive integers with
$d=2^{d_1}+2^{d_2}+\cdots +2^{d_s}$, $s\ge 2$ and $1\le
d_1<d_2<\cdots<d_s\le \lfloor{\rm log}_2n\rfloor$.
If $2d\not\preceq n$ or $2^{d_1+2}-1\preceq n$, then ${\rm
wt}(\sigma_{n, d})<2^{n-1}$.
\end{Corollary}

{\bf Proof}: According to Theorem \ref{Thm. 2^t weight}, ${\rm
wt}(\sigma_{n,2^t})\le 2^{n-1}$ if and only if $n\equiv r({\rm mod\
}2^{t+2})$, where $0\le r\le 2^{t+1}-1$ or $r=2^{t+2}-1$. That is
$2^{t+1}\not\preceq n$ or $2^{t+2}-1\preceq n$. Combined with
Corollary \ref{Cor. Weight}, if there exists $1\le i\le s$, such
that $2^{d_i+1}\not\preceq n$ or $2^{d_i+2}-1\preceq n$, then ${\rm
wt}(\sigma_{n, d})<2^{n-1}$. Since $2^{d_1+2}-1\preceq 2^{d_i+2}-1$
for any $1\le i\le s$, we have
\begin{center}$2^{d_i+2}-1\preceq n$ for some $1\le i\le s$
$\Longleftrightarrow$ $2^{d_1+2}-1\preceq n$.
\end{center}
Obviously, $2^{d_i+1}\not\preceq n$ for some $1\le i\le s$ iff
$2d\not\preceq n$.
This completes the proof .\hfill$\Box$

By the above results, we can obtain that ${\rm
wt}(\sigma_{n, d})<2^{n-1}$ holds in most cases, and so does the
conjecture.  

\begin{Remark}\label{Remark Cusick09 3.18}
Learned from the authors of \cite{Cusick09_JMC} we knew that the
result given in Lemma 3.18 \cite{Cusick09_JMC} holds for $0\le r\le
2^{t+1}$ instead for $0\le r<2^{t+1}+2^t$.


 If $d'$ is odd and $2^t+1<d'\le 2^{t+1}-1$ for some positive integer $t$.
Let $n'=2^{t+1}\cdot l'+r$. 
Since $2^t+1\preceq d'$ and $2^t+1\ne d'$, by Lemma \ref{Lemma ESB
Decomposition}, one has ${\rm wt}(\sigma_{n',
d'})<{\rm wt}(\sigma_{n', 2^t+1})$. Since 
\begin{eqnarray*}{\rm wt}(\sigma_{n', 2^t+1})=\sum_{s=0}^{2^{t-1}-1}
A_{n'}^{2^{t+1}}(2^t+2s+1)=2^{n'-2}+2^{n'-t-1}(-1)^{l'+1} \sum_{j=1,
odd}^{2^t-1}(\cos
\frac{j\pi}{2^{t+1}})^{n'-1}\frac{\sin\frac{jr\pi}{2^{t+1}}}{\sin
\frac{j\pi}{2^{t+1}}}.\end{eqnarray*} 

\begin{enumerate}
\item
If $r=0$ or $2^{t+1}$, then ${\rm wt}(\sigma_{n', 2^t+1})=2^{n'-2}$;

\item
If $1\le r< 2^{t+1}$, then $\sin(\frac{r\pi}{2^{t+1}})>0$. Since
$\frac{(\cos \frac{j\pi}{2^{t+1}})^{n-1}}{\sin\frac{j\pi}{2^{t+1}}}$
strictly decreases as $j$ increases for $1\le j\le 2^t-1$, by Lemma \ref{Lemma sum_T}, one
has ${\rm wt}(\sigma_{n', 2^t+1})<2^{n'-2}$ if $l'$ is even;

\item Similarly, if $l'$ is even and $2^{t+1}<r<2^{t+1}+2^t$, then
${\rm wt}(\sigma_{n', 2^t+1})>2^{n'-2}$.

\end{enumerate}

Thus, we can only obtain that ${\rm wt}(\sigma_{n', d'})\ne
2^{n'-2}$ for even $l'$ and $0\le r\le 2^{t+1}$.

\end{Remark}

\vspace{1mm}

\begin{Remark}\label{Remark Cusick08-09}
The above results cover the known results in \cite{Cusick08_IT,
Cusick09_JMC}.
\begin{enumerate}

\item It has been proved that
if $\sigma_{n, d}$ is balanced, then $d\le \lceil \frac{n}{2}\rceil$
\cite{Cusick08_IT}. In fact, 
if $wt(d)\ge 2$ and $\sigma_{n, d}$ is balanced,  by Corollaries
\ref{Cor. odd degree} and \ref{Cor. even degree}, one has $d$ must
be even and $2d\preceq n$. Thus, $d\le \lfloor \frac{n}{2}\rfloor$.

\item Let $n=2^{t+1}l-1$ for some positive integers $t$, $l$. If $d$ is even and
$2^t<d<2^{t+1}$, then $d$ can be written as
$d=2^{d_1}+2^{d_2}+\cdots+2^{d_{s-1}}+2^t$, where $s\ge 2$ and $1\le
d_1<d_2<\cdots<d_{s-1}<t$. Thus, $d_1+2\le t+1$ and
$2^{d_1+2}-1\preceq 2^{t+1}-1\preceq n$. By Corollary \ref{Cor. even
degree}, one has ${\rm wt}(\sigma_{n, d})<2^{n-1}$. So the result
obtained by Corollary 3.10 and Lemmas 3.1, 3.13 in
\cite{Cusick09_JMC} is a special case of Corollary \ref{Cor. even
degree};

\item
Let $n=2^{t+2}l+r-1$ for some positive integers $t$, $l$ and $0\le
r\le 2^{t+1}$. If $d$ is even and $2^t<d<2^{t+1}$, then $\sigma_{n,
d}$ is not balanced. The proof is as follows.
\begin{enumerate}

\item  If $r=0$, then $n=2^{t+2}l-1$. It is a special case of 2), and
so a special case of Corollary \ref{Cor. even degree};

\item  If $1\le r\le 2^{t+1}$. Since $d$ is even and $2^t<d<2^{t+1}$, $d=2^t+d'$
for some even integer $2\le d'\le 2^t-2$. Then,
$2d=2^{t+1}+2d'\not\preceq n$. By Corollary \ref{Cor. even degree},
one has ${\rm wt}(\sigma_{n, d})<2^{n-1}$.
\end{enumerate}
Thus, the result obtained by Corollary 3.10, Lemmas 3.1, and the
 modified Lemma 3.18 in \cite{Cusick09_JMC} (replace $0\le
r<2^{t+1}+2^t$ with $0\le r\le 2^{t+1}$) 
is also a special case of Corollary \ref{Cor. even degree}.
\end{enumerate}

\end{Remark}

%

\section{The Weight of $\sigma_{n,d}$}

In the section, we will discuss the weight of $\sigma_{n, d}$
depending whether $n\equiv 3({\rm mod\ }4)$ or $n\not\equiv
3({\rm mod\ }4)$.  If $n\equiv 3({\rm mod\ }4)$, our results cover
the results in \cite{Gao11_IT}. Furthermore, if $n=2^{t+1}l-1$,
$l\ge 3$ odd and $2^{t+1}|d$, then $\sigma_{n,d}$ is not balanced for
${\rm wt}(d)=1$ or $2d\not\preceq n$, which is not contained in
\cite{Gao11_IT}. We can also get  results for
$n\not\equiv 3({\rm mod\ }4)$. As a result, Conjecture \ref{Conjecture_08} can
be simplified to Conjecture \ref{Our Conjecture mod 4}.

\subsection{The Weight of $\sigma_{n,d}$ with $n\equiv 3({\rm mod\ }4)$}

When $n\ge 3$ and $n\equiv 3({\rm mod\ }4)$, it can be written as
$n=2^{t+1}l-1$, where $l\ge 1$ is odd and $t\ge 1$. For ${\rm
wt}(d)= 1$ and ${\rm wt}(d)\ge 2$, we can obtain the following
theorems, respectively. 
These results cover the results in \cite{Gao11_IT}.

\begin{Theorem}\label{Thm. 2^t 3 mod 4}
Let $n=2^{t+1}l-1$ and $d=2^s$, where $l\ge 1$ is odd, $t\ge 1$ and
$s\ge 1$. Then $\sigma_{n, d}$ is balanced if and only if $1\le s\le
t$.
\end{Theorem}

{\bf Proof}: By Corollary \ref{Cor. d=2^t}, $\sigma_{n, 2^s}$ is
balanced if and only if there exists $l'\ge 1$ such that
$n=2^{s+1}l'-1$. For any given positive integers $t$, $s$, and odd
$l$, there exists $l'\ge 1$ such that $2^{t+1}l-1=2^{s+1}l'-1$ if
and only if $1\le s\le t$. This finishes the proof. \hfill$\Box$

If $d>1$ is odd, by Corollary \ref{Cor. odd degree}, we have ${\rm
wt}(\sigma_{n,d})<2^{n-1}$.

If $d$ is even and ${\rm wt}(d)\ge 2$, we have the following
theorem.

\begin{Theorem}\label{Thm. 3 mod 4}
Let $n=2^{t+1}l-1$ and $2\le d=2^{t+1}d'+d''\le n$, where $l\ge 1$
is odd, $t\ge 1$, $d'\ge 0$ and $0\le d''<2^{t+1}$. If $d$ is even
and ${\rm wt}(d)\ge 2$, then ${\rm wt}(\sigma_{n,d})<2^{n-1}$ if
one of the following conditions hold:
\begin{enumerate}
\item  $l=1$;
\item  $l\ge 3$, $d''>0$;
\item  $l\ge 3$, $d''=0$, and $d'\not\preceq \frac{l-1}{2}$.
\end{enumerate}
\end{Theorem}

{\bf Proof}: Since $l$ is odd, write it as $l=2c+1$,
$c=\frac{l-1}{2}\ge 0$. Then $n=2^{t+1}l-1=2^{t+2}c+2^{t+1}-1$. If
$d$ is even and ${\rm wt}(d)\ge 2$, then $d$ can be written as
$d=2^{d_1}+2^{d_2}+\cdots+2^{d_s}$ with $s\ge 2$ and $1\le
d_1<d_2<\cdots<d_s\le \lfloor{\rm log}_2n\rfloor$.
\begin{enumerate}
\item If $l=1$, $n=2^{t+1}-1$.  Since $1\le d_1<d_s\le t$,
we have $d_1+2\le t+1$ and $2^{d_1+2}-1\preceq n$. By Corollary
\ref{Cor. even degree}, we obtain ${\rm wt}(\sigma_{n,d})<2^{n-1}$.

\item If $l\ge 3$ and $d''>0$, then $d''=2^{d_1}+2^{d_2}+\cdots+2^{d_i}$ for
some $1\le i\le s$. Since $d''<2^{t+1}$, one has $1\le d_1\le t$. By
Corollary \ref{Thm. 2^t 3 mod 4}, we get ${\rm wt}(\sigma_{n,
2^{d_1}})=2^{n-1}$. According to Corollary \ref{Cor. Weight},  ${\rm
wt}(\sigma_{n,d})<2^{n-1}$.

\item When $l\ge 3$ and $d''=0$.
If $d'\not\preceq c$, we have $2d=2^{t+2}d'\not\preceq n$. By
Corollary \ref{Cor. even degree}, we have ${\rm
wt}(\sigma_{n,d})<2^{n-1}$.
\end{enumerate}
The proof is completed. \hfill$\Box$

In \cite{Gao11_IT}, the authors proved that if $n=2^{t+1}l-1$, $l$
odd and $2^{t+1}\not |d$, $\sigma_{n,d}$ is balanced if and only if
$d=2^k$, $1\le k\le t$. 
Thus, the results in Corollary \ref{Cor. odd degree} and Theorems
\ref{Thm. 2^t 3 mod 4}, \ref{Thm. 3 mod 4}
cover the results given in \cite{Gao11_IT}. Furthermore, the
result for $l\ge 3$ odd and $2^{t+1}|d$ is not contained in
\cite{Gao11_IT}.


\subsection{The Weight of $\sigma_{n,d}$ with $n\not\equiv 3({\rm mod\ }4)$}

When $n\ge 3$ and $n\not\equiv 3({\rm mod\ }4)$, it can be written
as $n=2^{t+1}l+r$, where $l\ge 1$ is odd, $t\ge 1$, and $r=0$, $1$,
$2$. Similarly, we have the following results.


\begin{Theorem}\label{Thm. 2^t not 3}
Let $n\ge 3$ and $d=2^s\le n$ with $s\ge 1$. If $n\not\equiv 3({\rm mod\ }4)$, then
the elementary symmetric Boolean function $\sigma_{n, d}$ is not balanced.
\end{Theorem}

{\bf Proof}: The result directly follows from
 Corollary \ref{Cor. d=2^t}. \hfill$\Box$

If $d>1$ is odd, by Corollary \ref{Cor. odd degree}, we have ${\rm
wt}(\sigma_{n,d})<2^{n-1}$.

If $d$ is even and ${\rm wt}(d)\ge 2$, we have the following
result.

\begin{Theorem}\label{Thm. not 3 mod 4}
Let $n=2^{t+1}l+r$ and $2\le d=2^{t+1}d'+d''\le n$ with  $l\ge 1$ is
odd, $t\ge 1$, $r\in\{0, 1, 2\}$, $d'\ge 0$, and $0\le d''<2^{t+1}$.
If $d$ is even and ${\rm wt}(d)\ge 2$. Then, ${\rm
wt}(\sigma_{n,d})<2^{n-1}$ if one of the following conditions holds:
\begin{enumerate}
\item  $l=1$;
\item  $l\ge 3$, $d''=0$, and $d'\not\preceq \frac{l-1}{2}$;
\item  $l\ge 3$, $d''>0$, and ($d'\not\preceq \frac{l-1}{2}$ or
$d''\ne 2^t$).
\end{enumerate}
\end{Theorem}

{\bf Proof}: Since $l$ is odd, write it as $l=2c+1$,
$c=\frac{l-1}{2}\ge 0$. Then $n=2^{t+1}l+r=2^{t+2}c+2^{t+1}+r$.
\begin{enumerate}
\item When $l=1$, $n=2^{t+1}+r$.
Since $d$ is even, ${\rm wt}(d)\ge 2$, one has $4|2d$ and ${\rm
wt}(2d)\ge 2$. Thus, $2d\not\preceq n$. By Corollary \ref{Cor. even
degree}, we obtain ${\rm wt}(\sigma_{n,d})<2^{n-1}$.

\item When $l\ge 3$ and $d''=0$. If $d'\not\preceq c$, we have
$2^{t+2}d'\not\preceq 2^{t+2}c+2^{t+1}+r$. That is, $2d\not\preceq
n$. By Corollary \ref{Cor. even degree}, we obtain ${\rm
wt}(\sigma_{n,d})<2^{n-1}$.

\item When $l\ge 3$ and $d''>0$.
Since $d$ is even, one has $d''\ge 2$ and $4|2d''$. Therefore,
$2d''\not\equiv 2^{t+1}+1({\rm mod\ }2^{t+2})$ and $2d''\not\equiv
2^{t+1}+2({\rm mod\ }2^{t+2})$.

If $d''\ne 2^t$, then $2d''\not\equiv 2^{t+1}({\rm mod\ }2^{t+2})$.
So, $2d''\not\equiv 2^{t+1}, 2^{t+1}+1, 2^{t+1}+2({\rm mod\
}2^{t+2})$. From $2d''\ge 4>r$, one has $2d''\not\equiv r, 2^{t+1},
2^{t+1}+r({\rm mod\ }2^{t+2})$. Thus, $2d''\not\preceq 2^{t+1}+r$
and $2d\not\preceq n$. If $d'\not\preceq c$, we also have
$2d\not\preceq n$ . By Corollary \ref{Cor. even degree}, we get
 ${\rm wt}(\sigma_{n,d})<2^{n-1}$.
%
\end{enumerate}

The proof is completed.
\hfill$\Box$

From Corollary \ref{Cor. odd degree} and Theorems \ref{Thm. 2^t 3
mod 4}-\ref{Thm. not 3 mod 4}, Conjecture \ref{Conjecture_08} holds
in most cases and the unsolved conditions are
\begin{enumerate}
\item $d=2^{t+1}d'$,
$\textrm{wt}(d')\ge 2$ and $2\le d'\preceq \frac{l-1}{2}$ for
$r=-1$, $0$, $1$, $2$; Or
\item $d=2^{t+1}d'+2^t$,
$1\le d'\preceq \frac{l-1}{2}$ for $r=0$, $1$, $2$,
\end{enumerate}
where $n=2^{t+1}l+r$, $l\ge 3$ is odd, and $t\ge 1$.

Thus, Conjecture \ref{Conjecture_08} can be simplified as
Conjecture \ref{Our Conjecture mod 4}.

\section{The Weight of $\sigma_{n, 2^t+2^s}$}

In order to solve the conjecture, 
we consider the weight of $\sigma_{n, 2^t+2^s}$ and give some
experiment results on ${\rm wt}(\sigma_{n, 2^t+2^s})$ in this
section.



\begin{Theorem}\label{Thm. 2^t+2^s weight}
Let $n$, $d$ be two positive integers with $d=2^t+2^s\le n$ and
$1\le
t<s\le \lfloor{\rm log}_2n\rfloor$. Then 
\begin{eqnarray*}{\rm wt}(\sigma_{n,d})
&=&2^{n-2}-2^{n-s}\sum_{j=1, j\ odd}^{2^s-1}(\cos
a_j)^n\frac{\sin(2^ta_j)\sin((n-2^t+1)a_j)}{\sin(a_j)\sin(2^{t+1}a_j)}
.
\end{eqnarray*}
where $a_j=\frac{j\pi}{2^{s+1}}$.
\end{Theorem}

{\bf Proof}: Since $d\preceq i$ if and only if
$i=2^{s+1}i'+2^s+2^{t+1}p+2^t+q$ for some non-negative integers
$i'$, $p$, $q$ with $0\le p\le 2^{s-t-1}-1$ and $0\le q\le 2^t-1$.
Thus, $i\equiv 2^s+2^{t+1}p+2^t+q({\rm mod\ }2^{s+1})$ and
\begin{eqnarray*}{\rm wt}(\sigma_{n,d})&=&
\sum_{p=0}^{2^{s-t-1}-1}\sum_{q=0}^{2^t-1}A_n^{2^{s+1}}(2^s+2^{t+1}p+2^t+q)\\
&=&\sum_{p=0}^{2^{s-t-1}-1}\sum_{q=0}^{2^t-1}[2^{n-s-1}+2^{-s}\sum_{j=1}^{2^s-1}
(2\cos\frac{j\pi}{2^{s+1}})^n\cos\frac{j(n-2^{s+1}-2^{t+2}p-2^{t+1}-2q)\pi}{2^{s+1}}]
\\
&=&2^{n-2}+2^{n-s}\sum_{j=1}^{2^s-1}(\cos
a_j)^n\sum_{p=0}^{2^{s-t-1}-1}\sum_{q=0}^{2^t-1} \cos
((n-2^{s+1}-2^{t+2}p-2^{t+1}-2q)a_j)\\
&=&2^{n-2}+2^{n-s}\sum_{j=1}^{2^s-1}(\cos a_j)^n\cdot S_j,
\end{eqnarray*}
where $a_j=\frac{j\pi}{2^{s+1}}$ and
$S_j=\sum_{p=0}^{2^{s-t-1}-1}\sum_{q=0}^{2^t-1} \cos
((n-2^{s+1}-2^{t+2}p-2^{t+1}-2q)a_j)$ for $1\le j\le 2^s-1$. By
using the formula (\ref{eqn. cos(sx+y)}), we get
\begin{eqnarray*}S_j&=&\sum_{p=0}^{2^{s-t-1}-1}
\csc(-a_j)\cos((n-2^{s+1}-2^{t+2}p-2^{t+1}-2^t+1)a_j)\sin(-2^ta_j)\\
&=&\csc(a_j)\sin(2^ta_j)\sum_{p=0}^{2^{s-t-1}-1}
\cos((n-2^{s+1}-2^{t+2}p-2^{t+1}-2^t+1)a_j)\\
&=&\csc(a_j)\sin(2^ta_j)
\csc(2^{t+1}a_j)\cos((n-2^{s+1}-2^s-2^t+1)a_j)\sin(2^sa_j)\\
&=&\csc(a_j)\sin(2^ta_j)
\csc(2^{t+1}a_j)\cos((n-2^t+1)a_j-\frac{3j\pi}{2})\sin\frac{j\pi}{2}\\
&=&\left\{\begin{array}{lll}
0,\ \ &{\rm if\ } j\ {\rm is\ even},\\
-\csc(a_j)\sin(2^ta_j) \csc(2^{t+1}a_j)\sin((n-2^t+1)a_j), \ \ &{\rm
if\ } j\ {\rm is\ odd}.
\end{array}\right.
\end{eqnarray*}
Therefore,
\begin{eqnarray*}{\rm wt}(\sigma_{n,d})
&=&2^{n-2}-2^{n-s}\sum_{j=1, j\ odd}^{2^s-1}(\cos
a_j)^n\frac{\sin(2^ta_j)\sin((n-2^t+1)a_j)}{\sin(a_j)\sin(2^{t+1}a_j)}
.
\end{eqnarray*}
\hfill$\Box$

By Theorem \ref{Thm. 2^t+2^s weight}, we see that it is hard to
determine whether ${\rm wt}(\sigma_{n, 2^t+2^s})$ is greater than or
less than $2^{n-1}$. With the help of a computer, we calculate ${\rm
wt}(\sigma_{n, 2^t+2^s})$ and
 find that
\begin{enumerate}
\item if $t=1$ and $3\le l\le 181$, then ${\rm wt}(\sigma_{n, 2^t+2^s})<2^{n-1}$;

\item if $t=2$ and $l=3$, then $n=24+r$, ${\rm wt}(\sigma_{n, 12})>2^{n-1}$ and
${\rm wt}(\sigma_{n, 20})<2^{n-1}$;

\item if $t=2$ and $5\le l\le 121$, then ${\rm wt}(\sigma_{n, 2^t+2^s})<2^{n-1}$;

\item if $t\ge 3$, some of ${\rm wt}(\sigma_{n, 2^t+2^s})$ are greater
 than $2^{n-1}$;
\end{enumerate}
where $1\le
t<s\le \lfloor{\rm log}_2n\rfloor$,
$n=2^{t+1}l+r$, $l\ge 3$ is odd, and $r\in\{0, 1, 2\}$.

From Corollary \ref{Cor. Weight}, we have if ${\rm wt}(\sigma_{n,
2^t+2^s})<2^{n-1}$ and $2^t+2^s\preceq d$, then ${\rm wt}(\sigma_{n,
d})<2^{n-1}$.

\end{document}